\documentclass{IEEEtran} 
\usepackage{lipsum}
\usepackage{amsmath}
\usepackage{graphicx}
\usepackage[export]{adjustbox}
\usepackage{tikz}
\usepackage{pgfplots}
\usepackage{pgfplotstable}
\usepackage{amssymb}
\usepackage[caption=false]{subfig}
\usepackage[noend]{algpseudocode}
\usepackage[utf8]{inputenc}
\usetikzlibrary{decorations.pathreplacing}
\graphicspath{{images/}}
\usetikzlibrary{patterns}
\usepgfplotslibrary{fillbetween}
\pgfplotsset{compat=1.15,width=10cm}
\usepackage{cite}
\usepackage{optidef}
\usepackage{amsthm}
\usepackage[ruled,linesnumbered]{algorithm2e}

\usepackage{booktabs}
\usepackage{framed}
\usepackage{multirow}
\usepackage{float}
\makeatletter
\newcommand{\mathleft}{\@fleqntrue\@mathmargin0pt}
\newcommand{\mathcenter}{\@fleqnfalse}
\makeatother

\usepackage{comment}
\usepackage{float}

\title{A Lightweight Cell Switching and Traffic Offloading Scheme for Energy Optimization in Ultra-Dense Heterogeneous Networks}
\author{Attai Ibrahim Abubakar\IEEEauthorrefmark{1}, Michael S. Mollel\IEEEauthorrefmark{2}, Metin Ozturk\IEEEauthorrefmark{3}, Sajjad Hussain\IEEEauthorrefmark{1}, and Muhammad Ali Imran\IEEEauthorrefmark{1}
\thanks{\IEEEauthorrefmark{1}Communication, sensing and imaging (CSI) research group, James Watt School of Engineering, University of Glasgow, United Kingdom. 
Emails: a.abubakar.1@research.gla.ac.uk, 
sajjad.hussain@glasgow.ac.uk, and muhammad.imran@glasgow.ac.uk.
\IEEEauthorrefmark{2}Electrical and Electronics Engineering, Ankara Yıldırım Beyazıt University, Ankara, Turkey. Email: metinozturk@ybu.edu.tr.
\IEEEauthorrefmark{3}The Nelson Mandela African Institution of Science and Technology (NM-AIST), Arusha, Tanzania. Email: michaelm@nm-aist.ac.tz.}
}

\begin{document}
\maketitle
\begin{abstract}
One of the major capacity boosters for 5G networks is the deployment of ultra-dense heterogeneous networks~(UDHNs). However, this deployment results in tremendous increase in the energy consumption of the network due to the large number of base stations~(BSs) involved. In addition to enhanced capacity, 5G networks must also be energy efficient for it to be economically viable and environmentally friendly. Dynamic cell switching is a very common way of reducing the total energy consumption of the network but most of the proposed methods are computationally demanding which makes them unsuitable for application in ultra-dense network deployment with massive number of BSs. To tackle this problem, we propose a lightweight cell switching scheme also known as Threshold-based Hybrid cEll swItching Scheme~(THESIS) for energy optimization in UDHNs. The developed approach combines the benefits of clustering and exhaustive search~(ES) algorithm to produce a solution whose optimality is close to that of the ES~(which is guaranteed to be optimal), but is computationally more efficient than ES and as such can be applied for cell switching in real networks even when their dimension is large. The performance evaluation shows that the THESIS produces a significant reduction in the energy consumption of the UDHN and is able to reduce the complexity of finding a near-optimal solution from exponential to polynomial complexity. 

\end{abstract}
\section{Introduction}
The  proliferation of mobile phones, increasing use-cases of IoT devices, and the development of advanced mobile applications which are quite demanding in terms of bandwidth and latency, have led to increasing demand for mobile services. This has made mobile network operators~(MNOs) to continually increase their capacity through the deployment of more base stations~(BSs), thereby resulting in increased energy consumption~\cite{Adedoyin2020}. In addition, the introduction of network densification to cater for the projected 1000 times increase in capacity of 5G network compared to legacy networks would further heighten the energy consumption of the network~\cite{ALAMU2020}. From economic and environmental perspective, the aforementioned surge in capacity of 5G and beyond networks must not be done at the expense of huge energy consumption overhead. This is because increased energy consumption would result in more operational expenditure in form of increased electricity bills as well as environmental degradation due to increased $\text{CO}_2$ emission, as the energy used to power the BSs is mainly obtained from fossil fuels~\cite{usama2019survey, Buzzi2016}.

Various methods have been proposed for the optimization of energy consumption in ultra-dense networks~(UDN) including antenna muting~\cite{Frenger2021}, cell zooming~\cite{Ghosh2020}, power control~\cite{Aslani2020}, sectorization~\cite{ferhi2018}, dynamic cell switching operations~\cite{Abubakar2019}, etc.
Among all these methods, dynamic cell switching is the most commonly used approach because it results in the most energy savings, it is easier to implement and it requires minimal changes in the architecture of the network~\cite{ogbebor2020energy, Han2016, Feng2017}.
Several cell switching frameworks have been developed using machine learning and heuristic approaches. The machine learning techniques proposed for cell switching includes the use of supervised learning~(e.g., artificial neural networks~(ANN))~\cite{Abubakar2020}, unsupervised learning~(e.g., $k$-means)~\cite{dai2019cluster}, reinforcement learning~(e.g., multi-armed bandit, $Q$-learning)~\cite{Abubakar2019, Asad2019} and deep reinforcement learning~(e.g., deep and double-deep $Q$-learning)~\cite{Huang2020, kzhang2020}. 

Machine learning techniques are able to learn from historical and real time data to optimize network performance, have good generalization ability and are able to adapt to dynamic network conditions~\cite{Cayamcela2019}.
However, the challenge with most machine learning algorithms is that their training process is very computationally demanding and hence cannot be used for real-time network operation~\cite{ZHOU2017} A few machine learning algorithms such as $Q$-learning algorithms are quite computationally efficient when the network size is small. However, when the network size becomes very large, it becomes very challenging to apply as it involves learning a huge state-action table and also massive memory to store the learnt table~\cite{Hussain2020}.
Heuristic approaches~\cite{Luo2018, Dong2019}, on the other hand, are easier to implement, but have poor generalization ability and cannot adapt to dynamic network environment such as is obtained in 5G and beyond networks. As such, even though some of them are computationally efficient and can be applied to a large networks, they mostly result in sub-optimal solutions which could result in degradation in the quality of service~(QoS) of the network~\cite{Bui2017}.

As far as optimal cell switching solutions is concerned, the exhaustive search algorithm~(ES) also known as the brute-force algorithm always guarantees to find the optimal result because it sequentially searches all the possible combination of small BSs~(SBSs) and selects the best combination to switch off~\cite{Ozturk2021}. However, the computational complexity of the algorithm increases exponentially as the number of SBSs increases although it is computationally efficient and generates the optimal results very fast when the number of SBSs are few. An alternative approach could be obtained by compromising slightly on the optimality of ES while achieving a solution that is both scalable and of reduced complexity by first clustering the SBSs into smaller groups to reduce the search space, and then applying ES to each cluster separately in order to determine the set of SBSs to switch off per time.

Based on the aforementioned approach, we propose a cell switching framework known as Threshold-based Hybrid cEll swItching Scheme~(THESIS) that combines unsupervised machine learning scheme and ES algorithm for energy optimization in ultra-dense heterogeneous networks~(UDHN). The proposed method combines the advantages of unsupervised learning in terms of scalability and low complexity and ES algorithm in terms of optimality, to produce a cell switching strategy that is more computationally efficient but sub-optimal to the ES solution. In other words, the proposed approach tries to find a good trade-off between the performance and computational complexity, such that the computational complexity is significantly reduced without compromising much on the performance. In addition, the proposed framework can be applied even when the number of SBSs deployed in the network becomes very large, making it scalable and eliminating the limit of network size while applying the algorithm.

The rest of the paper is organized as follows: In Section II, a review of the related works is presented, while the system model is introduced in Section III. The proposed THESIS is presented and discussed in Section IV, while in Section V, the performance of the proposed scheme is evaluated. The paper is concluded in Section VI.

\section{related works}
There are various approaches in the literature for the implementation of cell switching in UDHN in order to minimize its energy optimization. In~\cite{Luo2019}, the authors considered the joint optimization of the area spectral efficiency~(ASE) and energy efficiency~(EE) of a two-tier UDHN. A firefly algorithm was developed to determine the optimal system parameters that would  jointly optimize the ASE and EE of the network. The authors in~\cite{Abdelradi2020} proposed an energy efficient traffic offloading framework for a heterogeneous network~(HetNet) based on queuing theory. A heuristic based traffic offloading algorithm was developed to maximize the EE of the network while ensuring that the stability of the queues is maintained.
Three heuristic algorithms were proposed in~\cite{habibi2021adaptive} for the maximization of the EE of a dense HetNet without compromising the QoS of users. 

The authors in~\cite{Luo2018} considered the problem of user association and dynamic SBSs switching for minimizing the energy consumption in UDNs while considering the switching energy cost. Two heuristic algorithms were proposed: the first is a centralized user association algorithm for minimizing the switching cost while the second is an enhanced heuristic for further reduction in the energy consumption of the network.
A cooperative energy optimization scheme for 5G UDNs using graph theory was proposed in~\cite{Daas2019}. The network was first represented as a graph after which the graph theory is employed to determine the switching off/on pattern of the SBSs in the network. 
The work in~\cite{Lee2021} proposed an energy efficient scheme for a two-tier HetNet via BS switching and traffic offloading. A distributed algorithm based on message-passing was developed to minimize the over-all power consumption of the network while maximizing the sum rate.
In~\cite{Dong2019}, the problem of power minimization in cached-nabled BSs was investigated while considering the limited resources and BS storage capacity. Three heuristic algorithms were developed to determine the sub-optimal bandwidth and user association strategy as well as minimize the power consumption of the network.

The challenge with heuristic algorithms is that they are hard-corded, have poor generalization ability and as such are not able adapt to dynamically changing network environment envisioned in 5G and beyond UDNs. Moreover, since the network conditions changes dynamically, there is a need for repeated application of the solution each time there is a change in the condition of the network, thereby resulting in huge time and computational overhead. Most times, before these computations are completed and the decision implemented, the network condition would have changed, thereby leading to sub-optimal results which would negatively affect the QoS of the network. Hence, they are not suitable for 5G UDHNs as they could lead to poor QoS and sub-optimal network performance~\cite{Bui2017, Hussain2020}.

The work in~\cite{Asad2019} considered the problem of energy consumption and $\text{CO}_2$ emission of a HetNet deployment in a smart city context. A $Q$-learning-based cell switching framework was proposed to reduce the energy consumption and $\text{CO}_2$ emission levels of the network. 
In~\cite{Asad2020}, a mobility management based energy optimization framework for HetNets was proposed using both supervised and $Q$-learning algorithms. The proposed framework uses supervised learning alongside historical data set of bus passengers passing through the HetNet to predict the traffic loads of the BSs while $Q$-learning was used to determine the cell switching and traffic offloading strategy that would minimize both the energy consumption and $\text{CO}_2$ emission level.
The work in~\cite{Amine2020} investigated the trade-off between energy and delay in a HetNet where the sleep mode of the SBSs can be adjusted to different levels for the purpose of energy saving while ensuring that the QoS is maintained. A distributed $Q$-learning algorithm was developed to adapt the sleep level of the SBSs to their activity level while considering co-channel interference.

A location-aware BS sleeping strategy that would jointly optimizes the trade-off between energy and delay in a 5G HetNet was introduced in~\cite{El-Amine2019}. A $Q$-learning algorithm which considers the location and velocity of the users in determining the sleep mode level of the BS was developed to maximize the energy delay trade-off of the network.
In~\cite{Panahi2018}, the authors proposed a cell switching mechanism using fuzzy $Q$-learning in order to maximize the energy efficiency of a HetNet without compromising the QoS. In addition, to avoid coverage holes when some BSs are switched off, a device-to-device communication mechanism was also incorporated into the sleeping mechanism.
The authors in~\cite{Zhang2020} proposed a dynamic framework for adjusting the load and energy consumption of the SBSs in a HetNet. The framework uses $Q$-learning to learn the optimal offloading policy required to turn off the redundant SBSs in the HetNet while balancing the load of the remaining SBSs.

$Q$-learning algorithms usually use tables~($Q$-tables) to store the learnt state-action values~($Q$-values). Hence, there is a $Q$-table entry for every action taken by the agent in the network environment. This approach is only feasible when the network dimension is small or medium. However, when the network dimension becomes very large, as obtained in 5G UDNs, it would becomes computationally burdensome to learn the $Q$-table, as the number of states or actions would greatly increase. In addition, a large memory would also be required to store the learnt $Q$ values. As a result, it is not feasible to use $Q$-learning for cell switching operation in UDNs~\cite{sutton2018reinforcement}

In~\cite{Abubakar2020}, an ANN based cell switching framework for energy optimization in UDN was proposed. The proposed framework is able to determine the optimal switching strategy that would lead to minimum energy consumption without violating the QoS of the network.
The authors in~\cite{Wang2017} proposed an online context-aware power optimization scheme for SBSs in cache-enabled HetNet. The energy minimization problem was first modelled as a multi-armed bandit problem then a Bayesian neural network was used to determine the optimal switching pattern that would optimize the energy consumption of the network.
A deep reinforcement learning and traffic prediction framework was designed in~\cite{QWu2021} for determining the sleeping strategy in a radio access network~(RAN). Their approach uses geographic and semantic spatial-temporal network~(GS-STN) for traffic forecasting while the BS sleeping problem was formulated as an MDP and solved using actor-critic reinforcement learning. 

The authors in~\cite{Huang2020} proposed an energy-aware traffic offloading scheme for EE optimization in HetNets. In the proposed scheme, the traffic demand of the network was first predicted using deep neural networks after which the traffic offloading strategy was obtained using deep $Q$-networks.
The work in~\cite{kzhang2020} developed a deep reinforcement learning framework for energy optimization in a RAN based on dynamic cell switch off/on. A double deep $Q$-learning network was developed to determine the optimal sleeping strategy that will minimize the energy consumption of the network while ensuring that the QoS of the network is maintained.
The authors in~\cite{Liu2018}, developed a dynamic BS sleeping strategy known as DeepNap. The proposed method employs deep $Q$-networks to learn the optimal sleeping policies of the BSs.
The challenge with neural network and deep reinforcement learning approaches is that their training process involves a very large computation overhead, which makes them unsuitable for real-time network operation.

A cluster-based femto BSs switching scheme to maximize the EE of a HetNet was developed in~\cite{huang2018cluster}, wherein semi-definite programming based correlation clustering algorithm was used to determine the cluster with minimum EE as well as the number of femto BSs to switch off within the cluster while considering load balancing and probability of outage. Similarly, in~\cite{dai2019cluster}, a cluster-based cell switching scheme for EE optimization in ultra dense SBS network was proposed while considering the user QoS and inter-cell interference. The EE problem was first formulated using stochastic geometry, then $k$-means algorithm was used to partition the dense SBSs into clusters. In addition, a sorting algorithm based heuristic, was developed to determine the number of SBSs to switch off in each cluster.
The authors in~\cite{li2018optimized} proposed clustering based sleeping strategy to minimize the power consumption and interference in dense HetNets using clustering algorithm. In their proposed method, the SBSs are clustered based on their interference level, after which the clusters with large interference values are selected. Then, a binary particle swarm optimization algorithm is applied to each of the selected clusters to determine the sleeping strategy.
Both clustering and sorting algorithms produce good results when the SBSs deployed are of the same type, however, as observed in~\cite{Ozturk2021}, their results become far from the optimality when different type of SBSs are deployed in the network.

In this paper, we propose THESIS, which leverages the optimality of ES algorithm and the low computational complexity of $k$-means clustering algorithm, to determine the switching strategy that would minimize the energy consumption of a UDHN. The proposed scheme can, without reasonable loss in optimality, determine the switching strategy of the SBSs with much lesser computational complexity compared to the optimal ES. It also has the advantage of being scalable and as such can be applied even when the number of SBSs becomes very large.
Different from previous works in~\cite{huang2018cluster, dai2019cluster, li2018optimized} where a single type of SBS is deployed, in this work, we consider a UDHN deployment where different types of SBSs~(remote radio head~(RRH), micro, pico, and femto BSs) are deployed under the coverage of the macro BSs~(MBSs). Moreover, unlike the previous works~\cite{huang2018cluster, dai2019cluster, li2018optimized} where a SBS is selected as the cluster head and is responsible for controlling the switching off/on of other SBSs, here the MBSs are responsible for controlling the switching off/on of the different SBSs because it has more global information about all the SBSs under its coverage.  
Also, we consider a scenario where horizontal traffic offloading is not always possible, as a result, vertical traffic offloading is employed in this work. In vertical traffic offloading, the traffic load of the SBSs that are turned off are transferred to the MBSs, in order to ensure the QoS of the network is maintained.

\subsection{Contributions}
A hybrid cells switching framework is developed to minimize the power consumption of a UDHN. The following are the contributions of this paper:
\begin{itemize}
    \item THESIS, a scalable cell switching approach based on $k$-means and ES algorithms is developed for energy optimization in UDHN. The proposed method is quite computationally efficient and produces results that are close to the optimal solution. It can also be applied to large scale networks where many SBSs are deployed.
    \item A benchmark algorithm purely based on $k$-means algorithm is developed for comparison with the proposed method.
    \item We also evaluate the quantity of $\text{CO}_2$ savings that can be obtained when the proposed cell switching approach is implemented. 
    \item A complexity comparison of the proposed algorithms with the benchmark algorithm is carried out in order to ascertain the computational efficiency of the proposed method.
    \item Finally, in order to capture the real life dynamics of the network, the performance of the proposed method is evaluated through extensive simulations using traffic data obtained from a real network and compared with other benchmark methods.
\end{itemize}

\section{system model}
\subsection{Network model}
\begin{figure}[t!]
\centering
\includegraphics[width=\columnwidth]{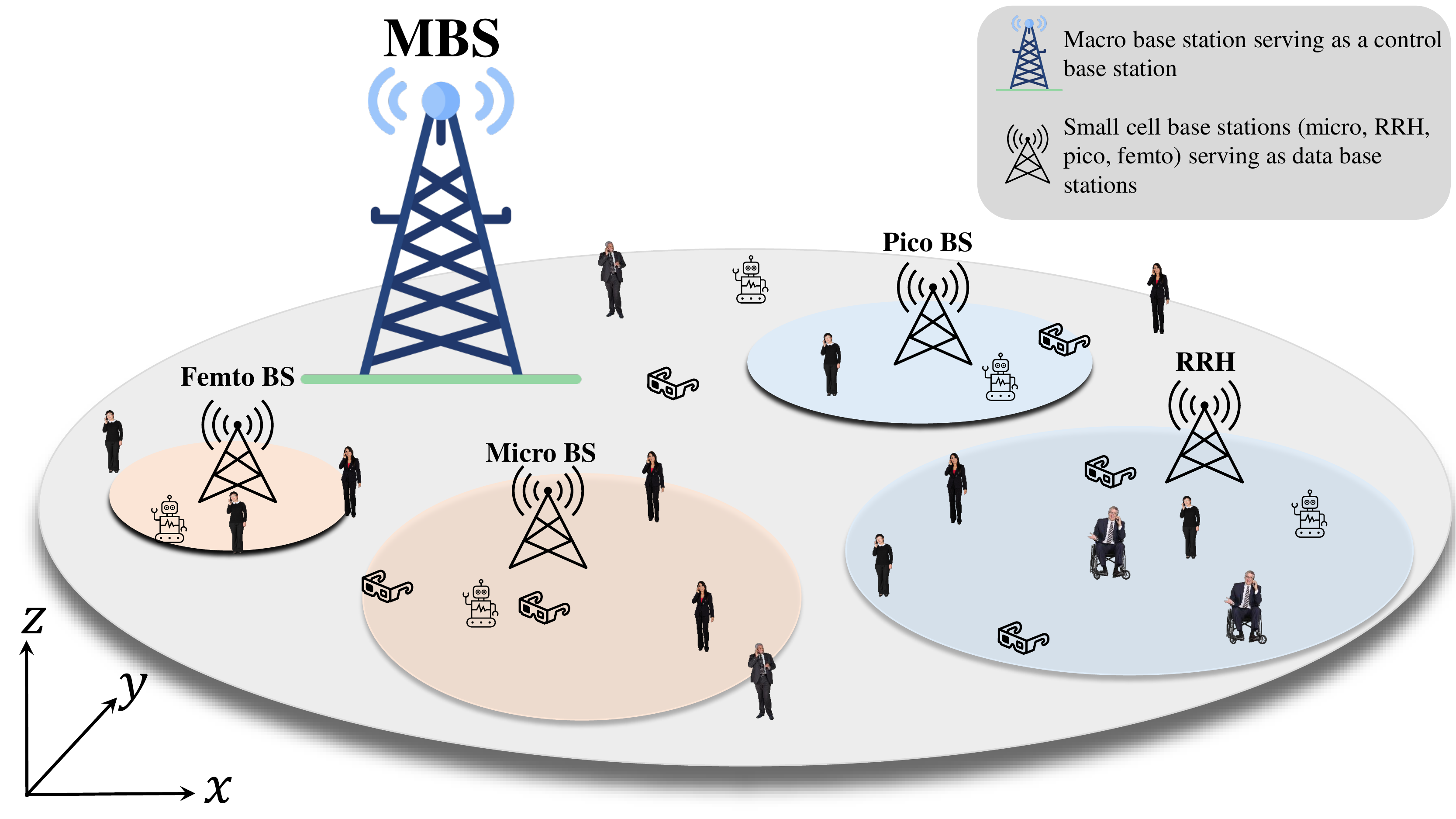}
\caption{An ultra-dense heterogeneous networks with one MC comprising a MBS and different types of SBSs~(RRH, micro, pico and femto BS).}\label{fig:sys_model}
\end{figure}

We consider a UDHN comprising multiple macro cells~(MCs). Each MC consists of a macro BS~(MBS) and a large number of SBSs each have different capacities and power consumption profiles. The UDHN employs control and data separated architecture~(CDSA) such that the MBSs serve as the control BS and are responsible for signalling, low data rate transmission and de~(activation) of SBSs under their coverage. The SBSs on the other hand serve as data BS, are deployed in areas with high traffic intensity for capacity enhancement and high data rate transmission. In addition, vertical traffic offloading is considered such that the SBSs with little or no traffic load can be turned off and the traffic originally associated with them is transferred to the MBS. This is to ensure that the QoS of the network is not violated. In this work we assume that before cell switching is implemented, all the traffic demands from the users are supported by the network, meaning that the UDHN always has sufficient radio resources for the users. In other words, because the UDHN is designed in terms of radio resources, when all the SBSs are on, all the users are guaranteed sufficient resources but with cell switching this cannot be guaranteed. Thus, we define the QoS to be the total amount of traffic demand that can be supported by the network after cell switching is implemented. The network model is presented in Fig.~\ref{fig:sys_model}. 

\subsection{Power Consumption of the UDHN}
The total power consumption of the UDHN consists of the power consumption of the MBSs and that of the SBSs. The power consumption of a BS,~$P_\text{BS}$, is given as~\cite{Auer2011, Debaillie2015}:
\begin{equation}\label{BS_pwr}
  P_\text{BS}~(\lambda, t) = P_\text{o} + \lambda_t \eta P_\text{tx},
\end{equation}
where $P_\text{o}$ is the constant circuit power consumption, $\lambda_t$ is the instantaneous traffic load, $\eta$ is the load dependent power consumption component and $P_\text{tx}$ is the transmission power of the BS. The value of $P_\text{o}$, $\eta$, and $P_\text{tx}$ depends on the type of SBS~(i.e., MBS, RRH, micro, pico and femto BS).

Therefore, considering a UDHN with multiple MCs, where the MCs are indexed by $i$, the total power consumption of the UDHN, $P_{\text{u}}$, is given by:
\begin{equation}\label{eqn:Pt1}
    P_{\text{u}}~(\lambda, t) =
    \sum_{i} \sum_{j} P_{\text{BS}_{i,j}}(\lambda, t),
\end{equation}
where $P_{\text{BS}_{i,j}}$ represents the power consumption of the $j^\text{th}$ BS in the $i^\text{th}$ MC, and $P_{\text{BS}_{i,1}}$ denotes power consumption of the MBS in the $i^{th}$ MC.

\section{Problem formulation}
We consider a certain duration of time~($T$) such that $T$ is partitioned into different time slots (in mins) of equal duration $L$ (in mins). Then, an index vector $u$ is defined which stores the time slots in a sequential order and can be expressed as $u=[1,2,...,K]$, where $K$ is the number of time slots and is written as $K=T/L$. We also consider a scenario where the UDHN can decide to switch off/on some SBSs during periods of low traffic in order to minimize the energy consumption of the network. Our goal is to determine the optimal switching strategy~(i.e., the optimal set of SBSs to turn off/on) in each time slot that would result in minimal energy consumption in the UDHN

Therefore, the total power consumption of the UDHN when cell switch off/on is considered can be expressed as: 
\begin{equation}\label{eqn:total_pwr}
   P_\text{u}~(\lambda, \Gamma_{i,j}) = \sum_{u} \sum_{i} \sum_{j} [\Gamma_{i, j} P_{\text{BS}i,j}(\lambda, t) + (1 - \Gamma_{i, j}) P^s_{\text{BS}i, j}],
\end{equation}
where $P^s_{\text{BS}i, j}$ denotes the sleep mode power consumption of the BS~(i.e., power consumption when switched off) and $\Gamma_{i, j}$ represents the off/on state of the $(i,j)^\text{th}$ BS at time $t$ i,e.,
    \begin{equation}\label{eq:on/off}
   \Gamma_{i, j}=
    \begin{dcases}
    1, &\text{if } B_{i,j} \,\, \text{is}\,\, \text{on}\\
    0, &\text{if } B_{i,j} \,\,\text{is}\,\, \text{off}
    \end{dcases}
    \end{equation}
Since the MBS is constantly active, $\Gamma_{i, 1} = 1$, $\forall t$.

The optimization objective is to minimize the total power consumption of the UDHN while ensuring that the quality of service~(QoS) of the network is maintained. Therefore, the power minimization objective function can be expressed as:
\begin{align}\label{eq:max_rev}
\underset{\Gamma_{i,j}}{\text{max.}}\quad
& \text{P}_{\text{u}}(\lambda, \Gamma_{i, j}),\\
\text{s.t.}\quad
& z =\hat \lambda_{i,1} + \sum_{j=2} \lambda_{i,j}\Gamma_{i,j}\label{eq:const_1},\\
& \hat\lambda_{i,1} \leq \lambda^\text{m}_{i,1}\label{eq:const_2},\\
& \Gamma_{i,j} \in \{0,1\}\label{eq:const_3}.
\end{align}
 
 
 
 

The constraint~\eqref{eq:const_1} is to ensure that the traffic demand that is supported by the UDHN before and after cell switching and traffic offloading is the same, where $z$ is the total traffic demand that was served by the UDHN before cell switching~(i.e., no traffic offloading was implemented), $ \hat \lambda_{i,1}$ is the traffic load of the MBS after traffic offloading and $\sum_{j=2} \lambda_{i,j}\Gamma_{i,j}$ is the total traffic of the remaining active SBSs. The second constraint given in~\eqref{eq:const_2} is to ensure the maximum traffic demand that can be supported by the MBS is not exceeded when offloading the traffic of the SBS that would be switched off to the MBS, where $\hat\lambda_{i,1}$ is the traffic load of the MBS after traffic offloading and $\lambda^\text{m}_{i,1}$ is the maximum traffic demand that can be served by the MBS.
The third constraint~\eqref{eq:const_3} denotes the off/on status of $(i,j)^\text{th}$ SBS as defined in~\eqref{eq:on/off}.

\section{Proposed Cell Switching Scheme}
The aim of this paper is to determine the optimal online policy for switching off/on the SBSs of the UDHN without compromising the QoS of the network. Popular heuristic approaches such as ES algorithm, even though always finds the optimal policy, due to huge computational complexity when the number of SBSs deployed becomes very large, is not suitable for this kind of problem. It is only suitable for application in networks with few number of SBSs as the optimal results in such cases are quicker to compute with the ES algorithm. Considering the limitation of applying ES algorithm particularly when the network size is very large, we propose a lightweight cell switching scheme known as THESIS, which combines $k$-means clustering and ES algorithms for energy optimization in UDHN. Before going into details about the proposed approach, for the sake of keeping the discussion easy to follow, we first introduce the benchmark scheme we developed known as multi-level clustering (MLC), which is purely based on $k$-means clustering algorithm. Therefore, in the following subsections, we will first discuss the foundations of cell clustering, followed by the benchmark MLC and the proposed approach, respectively.

\subsection{Cell clustering}
The basis for the development of both the benchmark and proposed cell switching algorithm is clustering SBSs with similar traffic load and deciding which cluster(s) or set of SBSs within a cluster can be switched off in order to minimize the total energy consumption of the UDHN. To cluster the SBSs, an unsupervised learning algorithm known as $k$-means algorithm is applied. However, the number of clusters must be determined in advance, before finding the members of each cluster, and thus the number of clusters becomes a hyper-parameter for the $k$-means algorithm. One possible solution to choose the optimal number of clusters is to use the elbow method~\cite{Yuan2019} Hence, in the following subsection, we briefly discuss the $k$-means algorithm followed by the elbow method.
\subsubsection{$k$-means algorithm}
The $k$-means algorithm is one of the clustering algorithms that is used to split an unlabelled data set into $k$ clusters, $C=\left\{C_{1}, C_{2}, \ldots, C_{k}\right\},$  where the optimal number of clusters, $k$, also represents the number of cluster centroids and $C_k$ denotes the $\textit{k}^\text{th}$ cluster.
The number of clusters is usually determined before hand using the elbow method~(which would be elaborated in the following paragraphs).
Hence, given the traffic loads of the SBSs in each MC of the UDHN, $\lambda_{i,j}$, and the optimal number of clusters, $k$ to partition $\lambda_{i,j}$, the task of the $k$-means algorithm is to minimize the intra-cluster distance between similar traffic loads and the centroid~(mean) of each cluster. The objective function of $k$-means algorithm, $J(k,\lambda, \mu_m)$, can be expressed as~\cite{Qin2017, JAIN2010KM}:

\begin{equation}\label{eqn:k-means}
\min_{\mu_m} J(k,\lambda, \mu_m) =\sum_{m=1}^{k} \sum_{\lambda_{i,j} \in C_{\mathrm{k}}}\left\|\lambda_{i,j} -\mu_m\right\|^{2},\end{equation}
where $\mu_m$ is the mean or center of the cluster $C_{\mathrm{k}}$. 

\subsubsection{Selection of optimal number of clusters~(elbow method)}
One of the most important aspects of clustering is determining the optimal number of clusters to split the data set. This is because the performance of the cluster-based cell switching algorithm depends on selecting the optimal number of clusters to group the SBSs in the UDHN.
The elbow method provides a suitable way of finding the optimal number of clusters from a giving data set. 
In the elbow method, the optimal number of clusters can be obtained by first evaluating the sum of the squares errors (SSE) between the data points in each cluster and the centroid to obtain $k$ values. The SSE can be expressed as~\cite{Yuan2019}:

\begin{equation}\label{eqn:elbow}
\mathrm{SSE}=\sum_{k=1}^{N} \sum \left(X - c_{k}\right)^{2}.
\end{equation}
where $k$ is the number of clusters, $X$ is the data points in a certain cluster and $c_k$ is the centroid of that cluster.
Then the SSE is plotted against the $k$ values. The value of $k$ where the SSE curve forms an elbow before flattening out is selected as the optimal number of clusters to be used in partitioning the data points in the data set.

\subsection{Multi-level clustering based cell switching scheme}\label{sec:MLC}
The MLC algorithm performs repeated clustering and re-clustering of the SBSs deployed within the coverage area of the MBS according to their traffic loads and attempts to offload the traffic load of the SBSs in the lightly loaded cluster to the MBS. Then the total power consumption of the UDHN after offloading the traffic of the cluster(s) to the MBS is determined. Finally, the cluster(s) which results in the least power consumption in the UDHN is selected as the optimal cluster(s).
The pseudo-code of the MLC algorithm is presented in Algorithm~\ref{algo:multilayeralgo} and the description of the algorithm is described as follows.

The elbow method is used to determine the optimum number of clusters. Based on the optimal number of clusters, $k$-means algorithm is applied to perform the first level of SBS clustering according to their traffic loads. After that, the aggregate traffic load of each cluster~($\lambda_{i,x}$) is computed and compared to the available radio resources at the MBS in order to determine the number of clusters that can be switched off. Then the power consumption of the network is computed, after offloading the traffic of each of the selected cluster(s) to the MBS.
The remaining clusters whose aggregated traffic load exceeds the maximum traffic demand that can be served by the MBS~($\lambda^\text{m}_{i,1} = 1$~(normalized)) are further divided into smaller clusters by repeating the preceding steps until only a single SBS is left whose traffic demand exceeds that of the MBS or all the SBSs have been exhausted and there are no more SBSs left.
Finally, the power consumption of the UDHN is computed after the various levels of clustering and traffic offloading have been carried out. The power consumption values obtained are then ranked in ascending order and the one with the lowest power consumption is selected as the optimal cluster/sub-cluster and all the SBSs in that cluster are switched off.
\begin{algorithm}[t!]
	\SetAlgoLined
	\SetKwInOut{Output}{output}
	\SetKwInOut{Input}{input}
	\Input{Traffic loads of MBS and SBSs}
	Initialize optimum number of cluster(s) to switch off, C$_\text{opt} \leftarrow 
	\text{None}$\;
	Initialize minimum energy saved by switching off cluster(s) $x$, E$_{{s}_\text{min}}$= 0\;
	Perform optimized $k$-means clustering with elbow-method to determine the optimal number of clusters, $k$\;
	Initialize the table, $\Psi$, containing the clusters and the traffic loads of the SBSs\;
	\For{$x$ $\in$ $k$}{
	\If{$\lambda_{i,x} + \lambda_{i,1} \le$ 1}
		{
				E$_{x}$ = Energy saved by switching off cluster $x$\;
				Remove cluster $x$ from $\Psi$\; 
				
				\If{E$_{x}$ $\ge$ E$_{{s}_\text{min}}$}{
				
				C$_\text{opt} \leftarrow x$\;
				E$_{{s}_\text{min}}$ $\leftarrow$ E$_{x}$
				
				}

		}
	
	}
	\uIf{there is any cluster left in $\Psi$}{
	Recursively return to step 3 by re-clustering each cluster left in $\Psi$;
	}
	\Else{
	\Output{C$_\text{opt}$}
	}
	\caption{MLC}
	\label{algo:multilayeralgo}
\end{algorithm}
\subsection{Threshold-based Hybrid Cell Switching Scheme}

\begin{algorithm}[t!]
	\SetAlgoLined
	\SetKwInOut{Output}{output}
	\SetKwInOut{Input}{input}
	\Input{Traffic loads of MBS and SBSs}
	Initialize B$_\text{th}$ as the maximum number of BS in the cluster.\;
	Initialize optimal SBS combination to switch off, BS$_\text{opt} \leftarrow \text{None}$\;
	Initialize minimum energy saved by switching off SBSs, E$_{{s}_\text{min}}$= 0\;
	Perform optimized $k$-means clustering with elbow-method to search optimal $k$-cluster.\;
	Initialize the table, $\Psi$, containing the clusters and the traffic load of the SBSs\;
	\For{$x$ $\in$ $k$}{
	\If {$\vert k_x \vert \le B_\text{th}$}{
	Run ES search and obtain the best combination of SBSs to switch off~(BS$_\text{cal}$) and their respective power consumption $E_{{BS}_\text{opt}}$\; 
	\If {$E_{{s}_\text{min}} \le E_{{BS}_\text{opt}}$ } 
	{${BS}_\text{opt} \leftarrow BS_\text{cal}$\;
	$E_{{s}_\text{min}} \leftarrow E_{{BS}_\text{opt}}$\;
	Remove cluster $k_x$ from $\Psi$\;
	}
	}

	}
	\uIf{there is any cluster left in $\Psi$}{
	Recursively return to step 3 by re-clustering individual cluster left in $\Psi$\;
	}
	\Else{
	\Output{BS$_\text{opt}$}
	}
	\caption{THESIS}
	\label{algo:multilayeralgothre}
\end{algorithm}

Even though the MLC method can be applied for cell switching when the network dimension is very large, however, it produces results that are very sub-optimal compared the ES algorithm. It should be noted that the goal of any sub-optimal algorithm is to produce a result that closely approximates the ES solution. As a result, we improve on the optimality of the MLC method by developing the THESIS algorithm, which combines the advantages of MLC in terms of scalability~(i.e., its applicability when the number of SBSs in the network becomes very large) and that of the ES algorithm in terms of optimality, to produce a solution that is close to the optimal result. In addition, the proposed THESIS is scalable and very computationally efficient compared to the ES method. The pseudo-code of the THESIS is presented in  algorithm.~\ref{algo:multilayeralgothre} while the procedure for the implementation of the algorithm is described as follows.

As in the MLC scheme, the elbow method is used to determine the optimum number of clusters.
Based on the optimal number of clusters, $k$-means algorithm is applied to perform the first level of clustering according to the traffic load of the SBSs.
Then, the clusters whose number of SBSs is less than the threshold value~($B_\text{th}$) are determined.
For those clusters with number of SBSs less than $B_\text{th}$, ES is applied to determine the number of SBSs to switch off and the power consumption of the network is computed.
For those clusters whose number of SBSs are greater than $B_\text{th}$, the $k$-means algorithm is applied to re-cluster the SBSs in those clusters after which ES is applied to determine the number of SBSs to switch off as well as the computation of power consumption of the network.
Finally, the power consumption of the UDHN is ranked based on the different sets of SBSs that were switched off. Then the set of SBSs that gives the least power consumption is selected as the optimal set of SBSs to switch off. 

The difference between the proposed THESIS and MLC is that MLC repeatedly clusters the SBSs and tries to find the cluster(s) to switch off based on the one that satisfies the constraints and also yields the minimum power consumption. On the other hand, THESIS goes a step further by searching within the clusters to select the set of SBSs that meet the constraints and gives a minimum power consumption. This ability of THESIS to search within the clusters enables it to discriminate between the different types of SBSs~(in terms of capacity and power consumption profile) when selecting the set of SBSs to switch off within the clusters while the MLC does not have this capability. Thus giving THESIS an advantage over the MLC.
\section{Performance Evaluation}
In this Section, the performance of the proposed scheme is evaluated using various metrics and compared with other benchmark algorithms. In addition, the complexity comparison of the proposed and benchmark methods is also carried out.
The proposed THESIS can be applied to any network irrespective of the network dimension in terms of number of MCs involved. It should be noted that since the UDHN comprises many MCs, each consisting of one MBS and several SBSs, the proposed framework is implemented at the MBS of each MC as it is responsible for controlling the operations of the SBSs within the MCs. Hence, only one MC in the UDHN is considered since the results obtained can be applied to all other MCS in the network.
The parameters used for the simulations are presented in Table 1.

\subsection{Traffic data and simulation parameters}
To compute the total power consumption of the UDHN using~\eqref{eqn:total_pwr}, the traffic load of the MBS and SBSs are required. The call detail record~(CDR) of the city of Milan that was made available by Telecom Italia~\cite{barlacchi2015multi} is used as the data set for this simulation. The data set has the city of Milan divided into 10000 square grids with each grid having an area of 235 $\times$ 235 square meters. The call, text message and internet activities performed in each grid was recorded with a 10 minutes resolution for two months~(November-December 2013) period. Even though the activity levels of the data set are unit-less and no information regarding how the data set was processed was provided, we have assumed that the CDR of each grid is their traffic load as they represent the amount of network resources utilized by the users within each grid for each time slot. However, in the course of data processing, we only considered the internet activity level as the traffic load for the BSs since we are investigating 5G networks which is mainly internet protocol based. The combination of the internet activity level of two randomly selected grids was used to denote the traffic load of the MBS while that of a single grid was considered for each of the SBSs. The traffic loads were then normalized with respect to the amount of radio resources of each of the SBSs in the UDHN~(i.e. RRH, micro, pico and femto cell). 

\begin{table}[t!]
\centering
\caption{Simulation Parameters}
\label{tab:Simulation Parameters}
\begin{tabular}{@{}ll@{}}
\toprule
\textbf{Parameter}                 & \textbf{Value}           \\ \midrule
Bandwidth of MBS                    & 20MHz                    \\ 
Bandwidth of SBSs, SN-BSs           & 15MHz, 10MHz, 5MHz, 3MHz \\ 
Number of RBs per MBS               & 100                      \\ 
Number of RBs per SBSs, SN-BSs      & 75, 50, 25, 15           \\ 
$P_\text{tx}$~(MBS)~(W) & 20\\
$P_\text{tx}$~(RRH, micro, pico, femto)~(W) & 20, 6.3, 0.13, 0.05  \\ 
$P_\text{o}$~(MBS)~(W) & 130\\
$P_\text{o}$~(RRH, micro, pico, femto) ~(W) & 84, 56, 6.8, 4.8    \\ 
$\eta$~(MBS, RRH, micro, pico, femto) & 4.7, 2.8, 2.6, 4.0, 8.0  \\ 
$P^\text{s}_{\text{BS}_{i,j}}$~(RRH, micro, pico, femto)(W)         & 56, 39, 4.3, 2.9         \\ 
$B_\text{th}$             & 12                        \\
$\zeta$             & 0.2556                     \\ 
\bottomrule
\end{tabular}
\end{table}

\subsection{Performance Metrics}
\begin{itemize}
    \item \textbf{Power Consumption:} This is the instantaneous power consumption of the UDHN during the simulation time for each method based on~\eqref{eqn:total_pwr}. This metric enables us to carefully evaluate the performance of each approach as it reflects the instantaneous changes in power consumption of the network at different times of the day.
    \item \textbf{Energy Saved:} This metric is used to quantify the total amount of energy~(in Joules) that is saved over the whole simulation time~(24 hours). The energy saved for the proposed and benchmark approaches are obtained by comparing the presented methods with the case where all the BSs (both MBS and SBSs) are always on (it will be referred to as all-always-on(AAO) hereafter), such that the energy consumption of the presented methods and AAO are determined, and the difference between the presented method and and AAO are individually calculated as their energy saved.  
 
    \item \textbf{Carbon Emission:} One of the benefits of energy optimization is that it ensures the reduction of the carbon foot print of the network. The carbon emission level of the network can be obtained from the total energy consumption with the help of the $\text{CO}_2$ conversion factor~($\zeta$). The $\text{CO}_2$ emission~($\mathcal{E}_{\text{CO}_2}$) associated with the energy consumption of the UDHN~($E_\text{u}$) can be expressed as~\cite{Asad2019}:
    \begin{equation}
       \mathcal{E}_{\text{CO}_2} = \zeta \sum_{t = 1} ^T E_{\text{u}, t}
    \end{equation}.
    \item \textbf{Average Network Throughput:} We consider the network throughput to be the effect that both the proposed and benchmark methods have on the QoS of the network after their implementation. In this work,  the network throughput is considered to be the traffic demand that can be supported by all active BSs~(both MBS and active SBSs) after cell switching operation has been implemented. In calculating the average throughput of the network, the throughput of the MBS and all active SBSs under its coverage is aggregated. This average network throughput can be expressed as:
    \begin{equation}
        \mathcal{T}_\text{N}(t) = \mathcal{T}_{i,1}(t) + \sum_{j=2} \mathcal{T}_{i,j}(t).
    \end{equation}
    where $\mathcal{T}_{i,j}(t)$ is the throughput of each type of SBS in the UDHN and $\mathcal{T}_{i,1}$ is the throughput of the MBS.
\end{itemize}

\subsection{Benchmarks}
\begin{enumerate}
    \item \textbf{ES:} This method yields optimum results and is always guaranteed to find the best switching pattern from all possible combination of SBSs switching patterns. It also considers the the amount of radio resources at the MBs when determining the best switching option such that the maximum traffic demand that can be served by the network is never exceeded. Hence, the QoS of the network is always guaranteed when this method is applied. The goal of any cell switching algorithm is to closely approximate this approach.
    \item \textbf{MLC:} This scheme has been described in detail in Section~\ref{sec:MLC}. It employs only $k$-means algorithm to determine the optimal number of clusters to switch off per time in order to minimize the total power consumption of the UDHN. This method involves much lesser computation overhead compared to ES, respects the QoS constraints as ES, and can be applied even when the network dimension is very large. However, Its performance is sub-optimal compared to the ES approach.
    
    \item \textbf{AAO:} In this approach, no cell switching is implemented, and as such all the SBSs are continuously left on. There is also no need for traffic offloading in this method because none of the SBSs are turned off. This method ensures that the QoS of the network is always maintained but there is no energy savings in this approach since the SBSs are always kept on. 
\end{enumerate}

It should be noted that tabular reinforcement learning approaches such as $Q$-learning, multi-armed bandit and deep reinforcement learning approaches such as deep and double-deep $Q$-networks that are known for intelligent decision making have not been considered as benchmarks in this work. This is because as pointed out in Sections I and II, it is computationally demanding to learn the state-action table and a large memory is required to store the learned state-action table when the network dimension is very large as considered in this work. It is also very computationally demanding to train deep reinforcement learning models~\cite{Hussain2020, sutton2018reinforcement} as such they cannot be applied for online cell switching operation considered in this work.

\subsection{Results and Discussions}
\begin{figure}[t!]
	\centering
	\label{fig:rev}
	\subfloat[$N = 20$]{%
		\resizebox{0.9\columnwidth}{!}{
%
%
\definecolor{mycolor1}{rgb}{0.00000,0.44700,0.74100}%
\definecolor{mycolor2}{rgb}{0.49400,0.18400,0.55600}%
\definecolor{mycolor3}{rgb}{0.85000,0.32500,0.09800}%
\begin{tikzpicture}

\begin{axis}[%
width=4.521in,
height=3.536in,
at={(0.758in,0.51in)},
scale only axis,
xmin=0,
xmax=23,
xtick={0,2,...,23},
xlabel style={font=\bfseries\color{white!15!black}},
xlabel={Time~(hr)},
ymin=5.600,
ymax=6.810,
ylabel style={font=\bfseries\color{white!15!black}},
ylabel={Power Consumption~(kW)},
grid=both,
axis background/.style={fill=white},
legend style={at={(0.995,0.0)}, anchor=south east, legend cell align=left, align=left, draw=white!15!black},
label style={font=\Large},
ticklabel style={font=\large},
legend style={font=\Large}
]

\addplot [color=black, line width=2.0pt, mark=o, mark options={solid, black}]
  table[row sep=crcr]{%
0	6.449557165\\
1	6.35677064\\
2	6.30574384\\
3	6.247989057\\
4	6.240298196\\
5	6.275748769\\
6	6.375484732\\
7	6.50517796\\
8	6.672618985\\
9	6.765319056\\
10	6.797119438\\
11	6.799730263\\
12	6.735160645\\
13	6.771775508\\
14	6.771554645\\
15	6.762373162\\
16	6.784709663\\
17	6.779797843\\
18	6.766523283\\
19	6.745776866\\
20	6.713874421\\
21	6.678391778\\
22	6.603660631\\
23	6.502823729\\
};
\addlegendentry{AAO}

\addplot [color=mycolor2, line width=2.0pt, mark=o, mark options={solid, mycolor2}]
  table[row sep=crcr]{%
0	6.231523922\\
1	6.124170223\\
2	6.051129172\\
3	5.98809233\\
4	5.982540442\\
5	6.017455899\\
6	6.14794048\\
7	6.30933555\\
8	6.511504488\\
9	6.634818267\\
10	6.679429665\\
11	6.673114783\\
12	6.596842687\\
13	6.646911509\\
14	6.649094903\\
15	6.645066322\\
16	6.668966838\\
17	6.659314835\\
18	6.645008761\\
19	6.609688236\\
20	6.577993768\\
21	6.525566053\\
22	6.434357013\\
23	6.314806922\\
};
\addlegendentry{MLC}

\addplot [color=mycolor1, line width=2.0pt, mark=square, mark options={solid, mycolor1}]
  table[row sep=crcr]{%
0	6.140249557\\
1	6.024602139\\
2	5.95156582\\
3	5.884036367\\
4	5.873184694\\
5	5.921725394\\
6	6.05503632\\
7	6.225988267\\
8	6.460122195\\
9	6.601912637\\
10	6.656455255\\
11	6.650572364\\
12	6.566400611\\
13	6.624716026\\
14	6.63246857\\
15	6.618431034\\
16	6.647714764\\
17	6.632265452\\
18	6.615848544\\
19	6.583500664\\
20	6.528704629\\
21	6.478293535\\
22	6.364808038\\
23	6.222463323\\
};
\addlegendentry{THESIS}

\addplot [color=mycolor3, line width=2.0pt, mark=asterisk, mark options={solid, mycolor3}]
  table[row sep=crcr]{%
0	6.027311935\\
1	5.870046129\\
2	5.787973226\\
3	5.683917742\\
4	5.672554194\\
5	5.744473871\\
6	5.915096452\\
7	6.130650645\\
8	6.411643548\\
9	6.57175129\\
10	6.630135484\\
11	6.629829032\\
12	6.536474839\\
13	6.598468387\\
14	6.603230968\\
15	6.591164516\\
16	6.625455161\\
17	6.607337419\\
18	6.591867419\\
19	6.550980645\\
20	6.496496452\\
21	6.436561935\\
22	6.306028065\\
23	6.123045484\\
};
\addlegendentry{ES}

\end{axis}
\end{tikzpicture}%
}\label{fig:P_20}}
		
	\subfloat[$N = 60$]{%
		\resizebox{0.9\columnwidth}{!}{
%
%
\definecolor{mycolor2}{rgb}{0.00000,0.44700,0.74100}%
\definecolor{mycolor1}{rgb}{0.49400,0.18400,0.55600}%
\begin{tikzpicture}

\begin{axis}[%
width=4.521in,
height=3.536in,
at={(0.758in,0.51in)},
scale only axis,
xmin=0,
xmax=23,
xtick={0,2,...,23},
xlabel style={font=\bfseries\color{white!15!black}},
xlabel={Time~(hr)},
ymin=16.500,
ymax=18.200,
ylabel style={font=\bfseries\color{white!15!black}},
ylabel={Power Consumption~(kW)},
grid=both,
axis background/.style={fill=white},
legend style={at={(0.76,0.0)}, anchor=south west, legend cell align=left, align=left, draw=white!15!black},
label style={font=\Large},
ticklabel style={font=\large},
legend style={font=\Large}
]

\addplot [color=black, line width=2.0pt, mark=o, mark options={solid, black}]
  table[row sep=crcr]{%
0	17.45765608\\
1	17.27499407\\
2	17.18235964\\
3	17.07749958\\
4	17.06133202\\
5	17.1328135\\
6	17.32519346\\
7	17.57481333\\
8	17.89861317\\
9	18.09540376\\
10	18.16214584\\
11	18.15453889\\
12	18.02379807\\
13	18.09886434\\
14	18.09357198\\
15	18.08804729\\
16	18.13000424\\
17	18.10983637\\
18	18.06718826\\
19	18.03456513\\
20	17.96842836\\
21	17.88200702\\
22	17.74692511\\
23	17.55115364\\
};
\addlegendentry{AAO}


\addplot [color=mycolor1, line width=2.0pt, mark=o, mark options={solid, mycolor1}]
  table[row sep=crcr]{%
0	17.18027027\\
1	16.9588144\\
2	16.84602447\\
3	16.71810477\\
4	16.71170669\\
5	16.80273693\\
6	17.01686335\\
7	17.3124718\\
8	17.69431898\\
9	17.92666564\\
10	18.00221024\\
11	17.98855282\\
12	17.84772836\\
13	17.93706354\\
14	17.92739651\\
15	17.92763161\\
16	17.97267547\\
17	17.94758012\\
18	17.90319749\\
19	17.86373838\\
20	17.79031645\\
21	17.69442881\\
22	17.52677314\\
23	17.29773061\\
};
\addlegendentry{MLC}

\addplot [color=mycolor2, line width=2.0pt, mark=square, mark options={solid, mycolor2}]
  table[row sep=crcr]{%
0	17.05714242\\
1	16.83852341\\
2	16.71966945\\
3	16.60096631\\
4	16.57046029\\
5	16.65782362\\
6	16.89018917\\
7	17.20266844\\
8	17.61948128\\
9	17.86793738\\
10	17.95699925\\
11	17.94833782\\
12	17.7932359\\
13	17.89530547\\
14	17.8929968\\
15	17.8840157\\
16	17.9358673\\
17	17.90294048\\
18	17.8498154\\
19	17.81546998\\
20	17.72934072\\
21	17.618652\\
22	17.44148852\\
23	17.18379102\\
};
\addlegendentry{THESIS}


\end{axis}
\end{tikzpicture}%
}\label{fig:P_60}}
		
	\caption{Instantaneous power consumption over a 24 hrs period for 20 and 60 SBSs.}\label{fig:P_total}
\end{figure}
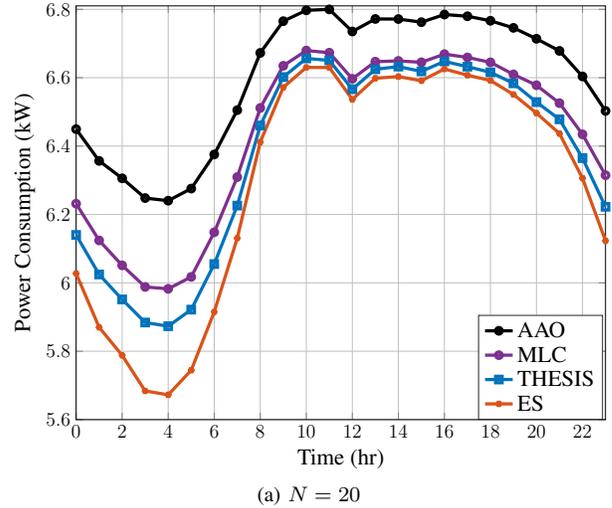
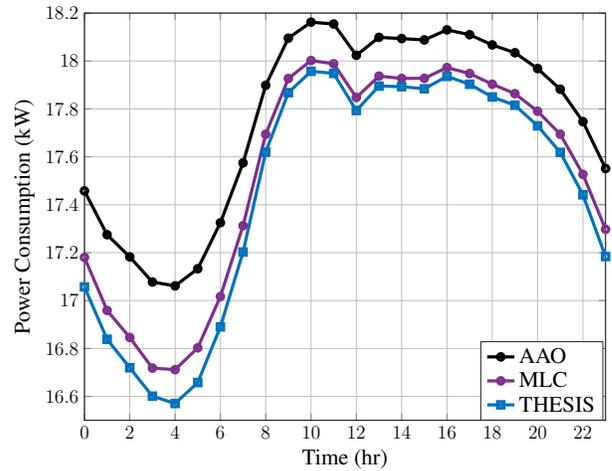

Fig.~\ref{fig:P_20} presents a comparison of the instantaneous power consumption of the proposed and benchmark methods over a 24 hour-period for 20 SBSs.
The first thing we can observe from  Fig.~\ref{fig:P_20} is that the pattern of power consumption of both the proposed and benchmark methods follow that of the traffic load of the network throughout the day such that it is low when the traffic load is low and high when the traffic load in high. The reason for this is that there are more opportunities to switch off many SBSs when the traffic load is low than when it is high, hence the discrepancies in power consumption values at different times of the day.
Second, the power consumption of the AAO method is higher than both the proposed THESIS and benchmark methods because no BS switching is performed in this method which means that all the SBSs are constantly kept on.
Third, the power consumption of the ES method is the lowest of all the methods including the proposed method because it searches sequentially through all the possible SBS switching combinations to select the option that leads to least energy consumption in the network at each time slot. However, this approach usually involves a huge computation overhead, which makes it to be only applicable to networks where the number of SBSs are few.

Fourth, we can observe that the performance of the proposed THESIS method is very close~(with 0.4\% difference) to that of the ES method when the traffic of the network is high but the margin becomes wider~(with 3.5\% difference) when the network traffic is low. The reason for the higher performance difference when the network traffic load low than when it is high is because during the periods of low traffic, even though there are more opportunities to switch off more SBSs, because the THESIS first partitions the SBSs into clusters before applying the ES to each cluster, the size of the search space becomes smaller. Hence, it has lesser opportunity to explore in order to determine the best switching combination that would result in lower power consumption in the network. On the other hand, during periods of high traffic load, the performance of the proposed method and the ES are much closer because there are very few opportunities to switch off the SBSs, therefore the higher search space of the ES does not give it much advantage over the proposed method.
However, when we analyze the time complexity of the proposed method and ES in Section~\ref{sec:6D_1}, it will be clear that the compromise in performance is greatly compensated with the complexity and scalability.

Fifth, it can be observed that apart from the AAO where no SBS is switched off, the performance of other methods exceeds that of MLC. This can be traced to the fact that MLC considers only the traffic loads of the SBSs when clustering and offloading the traffic of sleeping SBSs to the MBS, without considering that there are different types of SBSs~(with different capacity and power consumption profiles), and as such it might just be preferable to switch off a few SBSs with higher power consumption than many SBSs with low power consumption. It can also be observed that the performance of the MLC closely follows that of the proposed THESIS particularly during periods of high network traffic, however, the THESIS is able to outperform MLC more during periods of low traffic load because in addition to clustering the SBSs according their traffic load, ES is also applied to each cluster which enables it to discriminate among the different types of SBSs in order to select the best combinations of SBSs that would result in lesser power consumption in the network compared to the MLC approach. However, during periods of high traffic load in the network, the difference in power consumption between both methods is not significant because there are very few opportunities to switch off SBSs and as such lesser opportunity to search within the clusters so their performance becomes very close during such periods.   

Fig.~\ref{fig:P_60} shows the power consumption of the UDHN when 60 SBSs are deployed. It should be noted that ES algorithm is not considered in the scenario due to the huge computation overhead involved as well as limitations in our computing capacity. It can be observed that the trend of the power consumption when the proposed THESIS and benchmark methods are applied in Fig.~\ref{fig:P_60} follows the traffic load of the network as obtained in Fig.~\ref{fig:P_20} except that the magnitude of power consumption of the network is much higher in Fig.~\ref{fig:P_60} because more SBSs are deployed in this scenario compared to the previous one.
This finding is also quite intuitive, since the total power consumption of the network ($P_{\text{u}}$) is the cumulative sum of the power consumption of all the BSs involved in the networks, as seen in~\eqref{eqn:Pt1}, and thus once the network dimension rises, the total power consumption also increases.
The power consumption of the AAO method is also higher in this scenario compared to all other methods because no SBSs is turned off but are all left on to serve user demands.
The performance of MLC also follows that of the proposed THESIS in this scenario with the THESIS performing much better than MLC during periods of low traffic. This is due to the superior ability of THESIS to determine the best set of SBSs from the the various clusters to switch off that would result in lesser power consumption in the network rather than trying to switch off a whole cluster without considering the types of SBS in the cluster as this affects the magnitude of power consumption that can be obtained.

\begin{figure}[t!]
	\centering
	{\resizebox{0.9\columnwidth}{!}{
%
%
\definecolor{mycolor2}{rgb}{0.00000,0.44700,0.74100}%
\definecolor{mycolor1}{rgb}{0.49400,0.18400,0.55600}%
\definecolor{mycolor3}{rgb}{0.85000,0.32500,0.09800}%
\begin{tikzpicture}

\begin{axis}[%
width=4.521in,
height=3.536in,
at={(0.758in,0.51in)},
scale only axis,
xmin=8,
xmax=120,
xtick={  8,  16,  24,  36,  48,  60,  72,  84,  96, 108, 120},
xlabel style={font=\bfseries\color{white!15!black}},
xlabel={Number of SBSs},
ymin=20.000,
ymax=85.000,
ylabel style={font=\bfseries\color{white!15!black}},
ylabel={Energy Saved~(kJ)},
grid=both,
axis background/.style={fill=white},
legend style={at={(0.760,0.0)}, anchor=south west, legend cell align=left, align=left, draw=white!15!black},
label style={font=\Large},
ticklabel style={font=\large},
legend style={font=\Large}
]

\addplot [color=mycolor3, line width=2.0pt, mark=asterisk, mark options={solid, mycolor3}]
  table[row sep=crcr]{%
8	41.481\\
12	54.075\\
16	59.092\\
20	64.882\\
};
\addlegendentry{ES}

\addplot [color=mycolor2, line width=2.0pt, mark=square, mark options={solid, mycolor2}]
  table[row sep=crcr]{%
8	33.680\\
12	42.130\\
16	50.120\\
20	54.460\\
24	58.760\\
36	64.820\\
48	68.170\\
60	73.180\\
72	75.230\\
84	76.690\\
96	77.060\\
108	78.650\\
120	80.170\\
};
\addlegendentry{THESIS}

\addplot [color=mycolor1, line width=2.0pt, mark=o, mark options={solid, mycolor1}]
  table[row sep=crcr]{%
8	28.930\\
12	33.720\\
16	38.440\\
20	40.830\\
24	43.810\\
36	47.990\\
48	51.210\\
60	53.970\\
72	56.010\\
84	56.980\\
96	58.070\\
108	59.300\\
120	61.250\\
};
\addlegendentry{MLC}

\node[anchor=north west, red, align=center] at(axis cs:30,40) (source) {\large ES is terminated due \\ \large to computational burden};
\draw[-latex, ultra thick, red, dashed, shorten >= 1.5em](source) -- (axis cs:18, 68.00);

\end{axis}
\end{tikzpicture}%
}}
	\caption{Energy saved in the UDHN for different number of SBSs over a 24 hrs period.}
	\label{fig:energy120}
\end{figure}
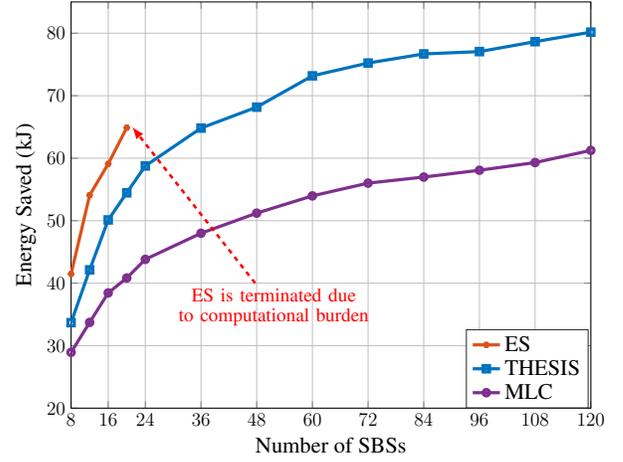

Fig.~\ref{fig:energy120} presents the total energy saved over a 24 hour period when the proposed and benchmark methods are applied for different number of SBS deployment. The first thing that can be observed from Fig.~\ref{fig:energy120} is that the total energy saved in the network increases as the number of SBSs deployed increases. This is due to the fact that with more SBS deployment, there are more opportunities to switch off many SBSs which leads to more energy saving. 
It can also be observed that the ES method gives the highest energy saving of all of the methods applied. However, it is accompanied by very high computation overhead and hence it cannot be applied in real networks with very large dimension. As a result, we had to stop the simulation at 20 SBSs due to the limitations of our device to handle such computation complexity.

The magnitude of the energy saved in the network when the proposed THESIS is applied also increases as the number of SBSs deployed increases. It can also be observed that the energy saved increases with high magnitude as the number of SBSs increases until when the number of SBSs reaches 60, afterwards the difference in energy saving between successive SBS deployments becomes smaller and almost constant. The rationale behind this is that even though there are more opportunities to switch off more SBSs as the 
number of SBSs increases, due to limited amount of radio resources at the MBS, the difference in the amount of SBSs that can be switched off is not much after 60 SBSs. 
The energy saving performance of the THESIS is quite lesser than that of the ES because of the wider search space that is available for searching for the optimal solution in the ES compared to the THESIS, however, the much lesser computation overhead involved in former compared to the latter makes it a more preferable for practical network deployment comprising many SBSs. 

The energy savings of the MLC method also increases with the number of SBS deployment, however, its energy saving seems to flatten out faster than the THESIS approach. The inability of the MLC to discriminate between the different types of SBSs when clustering the SBSs accounts for its lesser performance compared to the proposed method while the limitation in the amount of radio resources needed for offloading the traffic SBSs at the MBS accounts for the lesser difference in energy saving in the MLC method after 60 SBSs similar to what is observed in the THESIS method. 
Finally, we can observe that the difference in the energy saving between the THESIS and MLC also increases with higher magnitude as the number of SBSs increases until about 60 SBSs when it becomes almost constant. The reason is that there are more opportunities to switch off more SBS as the number of SBSs increases which accounts for more energy savings in both THESIS and MLC while the THESIS is able to discriminate among the different types of SBSs when making a cell switching decision thus making it produce higher energy saving compared to MLC. However, the almost constant energy saving difference observed after about 60 SBS is due to insufficient radio resources at the MBS to accommodate more traffic from the SBSs before turning them off. 

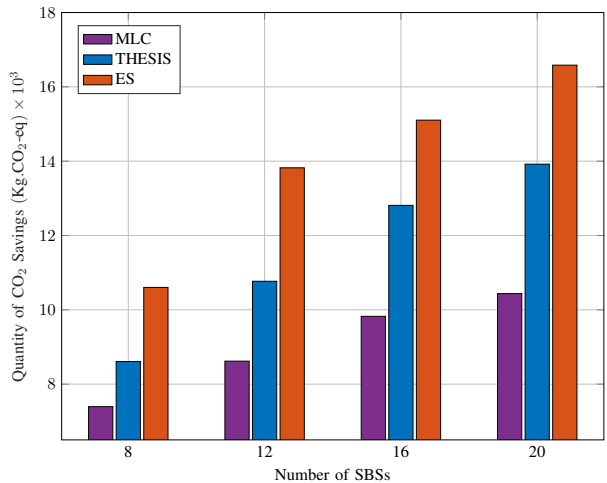
\begin{figure}[t!]
	\centering
	{\resizebox{0.9\columnwidth}{!}{
%
%
\definecolor{mycolor1}{rgb}{0.00000,0.44700,0.74100}%
\definecolor{mycolor2}{rgb}{0.85000,0.32500,0.09800}%
\definecolor{mycolor3}{rgb}{0.49400,0.18400,0.55600}%
\begin{tikzpicture}

\begin{axis}[%
width=4.521in,
height=3.566in,
at={(0.758in,0.481in)},
scale only axis,
bar shift auto,
xmin=0.511111111111111,
xmax=4.48888888888889,
xtick={1,2,3,4},
xticklabels={{8},{12},{16},{20}},
xlabel=Number of SBSs,
ymin=6.500,
ymax=18.000,
ylabel style={font=\color{white!15!black}},
ylabel={Quantity of $\text{CO}_2$ Savings $(\text{Kg}.{\text{CO}_2}\text{-eq}) \times 10^3$},
axis background/.style={fill=white},
xmajorgrids,
xminorgrids,
ymajorgrids,
yminorgrids,
legend style={at={(0.03,0.97)}, anchor=north west, legend cell align=left, align=left, draw=white!15!black}
]
\addplot[ybar, bar width=0.178, fill=mycolor3, draw=black, area legend] table[row sep=crcr] {%
1	7.394508\\
2	8.618832\\
3	9.825264\\
4	10.436148\\
};
\addplot[forget plot, color=white!15!black] table[row sep=crcr] {%
0.511111111111111	0\\
4.48888888888889	0\\
};
\addlegendentry{MLC}

\addplot[ybar, bar width=0.178, fill=mycolor1, draw=black, area legend] table[row sep=crcr] {%
1	8.608608\\
2	10.768428\\
3	12.810672\\
4	13.919976\\
};
\addplot[forget plot, color=white!15!black] table[row sep=crcr] {%
0.511111111111111	0\\
4.48888888888889	0\\
};
\addlegendentry{THESIS}

\addplot[ybar, bar width=0.178, fill=mycolor2, draw=black, area legend] table[row sep=crcr] {%
1	10.6025436\\
2	13.82157\\
3	15.1039152\\
4	16.5838392\\
};
\addplot[forget plot, color=white!15!black] table[row sep=crcr] {%
0.511111111111111	0\\
4.48888888888889	0\\
};
\addlegendentry{ES}

\end{axis}
\end{tikzpicture}%
}}
	\caption{Quantity of $\text{CO}_2$ saved for different number of SBSs over a 24 hour period.}
	\label{fig:barCO2}
\end{figure}

Fig.~\ref{fig:barCO2} presents the quantity of $\text{CO}_2$ saving that is obtained when the proposed and benchmark methods are applied to the UDHN with different number of SBSs. Note that the $\text{CO}_2$ saving up to 20 SBSs is shown here so as to study the relative performance of the all the algorithms since the ES approach cannot be applied beyond this number of SBS due to computation complexity. It is important for us to quantify amount of $\text{CO}_2$ savings because one of the goals of the proposed cell switching algorithm is to ensure that the carbon foot print or quantity of $\text{CO}_2$ emission associated with the UDHN is greatly reduced by reducing the amount of energy consumption of the network as most of the energy used to power the BSs are obtained from fossil fuels. Thus, a reduction in the energy consumption in the network leads to a reduction in amount of energy demanded which translates in lesser $\text{CO}_2$ emission thereby resulting in environmental conservation and prevention of global warming~\cite{Buzzi2016,Mughees2020}.

From Fig.~\ref{fig:barCO2}, it can be observed that the quantity of $\text{CO}_2$ saving increases as the number of SBSs increases because there are more opportunities to switch off more SBSs, which translates to greater $\text{CO}_2$ saving. The ES algorithm gives the highest $\text{CO}_2$ saving but as already observed previously, it computational complexity limits its application in large scale networks such as UDHN. The $\text{CO}_2$ saving of the proposed THESIS algorithm is about 18\% lesser than that of ES due to the better switching ability of ES, however its computation efficiency makes it more suitable for application in real network even when their dimension is very large.
The MLC approach produces the least $\text{CO}_2$ savings because of its sub-optimal performance compared to the proposed method even though it is most computationally efficient, its very sub-optimal performance does not make it suitable for application in large networks.

The average throughput metric has been considered a measure of the QoS of the network in order to ensure that the constraint in~\eqref{eq:const_1} and~\eqref{eq:const_2} are maintained. The QoS of the network is maintained by ensuring that the traffic demand that is served by the network before and after cell switching is performed remains constant by offloading the traffic of the SBSs to the MBS and ensuring that the maximum traffic demand that can be served by the MBS is not exceeded during traffic offloading.
Both the proposed THESIS and the benchmark methods are able to ensure that the QoS of the UDHN is not violated. 
THESIS is carefully designed such that it checks whether a given combination of SBSs in each cluster can be offloaded to the MBS before proceeding to switch them off. A similar approach is also employed in the ES approach in order to ensure that the capacity of the MBS is not exceeded. For the MLC approach, the aggregate traffic of each cluster is also compared with the available radio resources at the MBS to see whether it can accommodate it before turning off the cluster(s). 

\subsubsection{Complexity}\label{sec:6D_1}
\begin{figure}[t!]
	\centering
	{\resizebox{0.9\columnwidth}{!}{\definecolor{mycolor0}{rgb}{0,0,0}%
\definecolor{mycolor2}{rgb}{0.00000,0.44700,0.74100}%
\definecolor{mycolor1}{rgb}{0.49400,0.18400,0.55600}%
\definecolor{mycolor3}{rgb}{0.85000,0.32500,0.09800}%
\begin{tikzpicture}
\pgfplotsset{
    scale only axis,
    xmin=0, xmax=120,
    xtick={  8,  16,  24,  36,  48,  60,  72,  84,  96, 108, 120},
      label style={font=\Large},
ticklabel style={font=\large},
}
\pgfplotsset{every axis y label/.append style={mycolor0},
             every y tick label/.append style={mycolor0},
             y axis  line style={mycolor0}}
\begin{axis}[
  axis y line*=right,
  grid=both,
  ymin=0, ymax=3500,
  xlabel=Number of SBSs,
  ylabel= Time~(secs),
label style={font=\Large},
ticklabel style={font=\large},
]
\addplot [color=mycolor1, line width=2.0pt, mark=o, mark options={solid, mycolor1}]
  coordinates{
    (8,69)
    (12,101.666666666667)
    (16,126.400000000000)
    (20,172.000000000000)
    (24,235.200000000000)
    (36, 312.400000000000)
    (48, 415.200000000000)
    (60, 522.400000000000)
    (72, 622.600000000000)
    (84, 829)
    (96, 1037.60000000000)
    (108, 1270.33333333333)
    (120, 1569)
}; \label{plot_one}

\addplot [color=mycolor2, line width=2.0pt, mark=square, mark options={solid, mycolor2}]
  coordinates{
    (8, 53)
    (12, 105.666666666667)
    (16, 188.000000000000)
    (20, 296.400000000000)
    (24, 456)
    (36, 660)
    (48, 904.600000000000)
    (60, 1229.40000000000)
    (72, 1583.80000000000)
    (84, 1980.00000000000)
    (96, 2429)
    (108, 2891.66666666667)
    (120, 3447)
}; \label{plot_two}

\end{axis}

\pgfplotsset{every axis y label/.append style={mycolor3},
             every y tick label/.append style={mycolor3},
             y axis  line style={mycolor3}}
\begin{axis}[
  axis y line*=left,
  axis x line=none,
  ymin=0, ymax=4500,
  ytick={  0,  643,  1286,  1929,  2560,  3200, 3840, 4480},
  ylabel= Time~(secs),
  legend style={at={(0.72,0.001)}, anchor=south west, legend cell align=left, align=left, draw=white!15!black},
  label style={font=\Large},
ticklabel style={font=\large},
legend style={font=\Large}
]
\addplot [color=mycolor3, line width=2.0pt, mark=asterisk, mark options={solid, mycolor3}]
  coordinates{
    (8, 0.174529000000000)
    (12, 12.1759243333333)
    (16, 986.290883465058)
    (20, 1643.04363010843)
    (24, 4399.08285432529)
}; \addlegendentry{ES}
\addlegendimage{/pgfplots/refstyle=plot_two}\addlegendentry{THESIS}
\addlegendimage{/pgfplots/refstyle=plot_one}\addlegendentry{MLC}

\end{axis}

\end{tikzpicture}}}
	\caption{Time complexity: total time taken to complete the simulation for different number of SBSs.}
	\label{fig:time}
\end{figure}
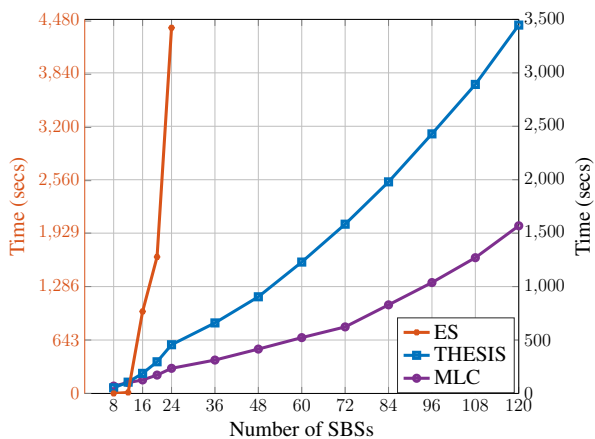

One of the ways of evaluating the complexity of an algorithm is to determine its time complexity,  that is, the simulation run time or time taken for the simulation to be complete~\cite{Liang2020}.
Fig.~\ref{fig:time} presents the time complexity comparison between the proposed THESIS approach and the benchmark methods. 
From Fig.~\ref{fig:time}, it can be observed that with ES algorithm, when the number of SBSs are few~(i.e., less than 16), the time complexity is very low, but from 16 SBSs and above, there is an exponential rise in the computational complexity. This is because the number of search spaces increases exponentially with every increment in the number of SBSs. Therefore, even though ES is always guaranteed to give the optimal switching strategy, due to its huge computational overhead,  it is not feasible to apply it for cell switching in UDHN comprising large number of SBSs.

The time complexity of the MLC algorithm can be observed to increase gradually and almost linearly. The complexity of MLC is very low because the number of clusters formed do not greatly increase as the number of SBSs increases, hence lesser time is required to select which cluster to switch off. Though the MLC is the most computationally efficient method, it is the least optimal approach and may not lead to much energy saving in the UDHN when applied.

The time complexity of THESIS is also quite low when the number of SBSs are less than 20, but afterwards, its time complexity begins to increase with a higher magnitude compared to the MLC. Overall, THESIS exhibits a polynomial time complexity which is because in addition to clustering, it also involves searching for the optimal combination of SBSs to switch off in each cluster. However, with the introduction of $B_\text{th}$, which limits the number of SBSs in the clusters where ES would be applied, the number of search spaces is reduced. The time complexity of THESIS is much lesser than that of ES and slightly higher than that of MLC but its energy saving performance is much closer to the optimal solution than the MLC. Therefore, THESIS is most suitable for application in UDHN compared to the ES and MLC.

\section{Conclusions}
In this work, we considered the problem of energy minimization in UDHN and proposed a computationally efficient cell switching scheme known as THESIS to determine the switching strategy that would result in the minimum energy consumption for the network. The proposed method is scalable and is able to---without much loss in optimality---produce a solution that has much lesser computation complexity compared to ES algorithm and can be applied to a network where a large number of SBSs are deployed. The performance evaluation shows that the proposed method is able to produce a significant reduction in the energy consumption compared to AAO as well as a decrease in the $\text{CO}_2$ emission level of the network with less complexity. A benchmark cell switching scheme using MLC was also developed which though is more computationally efficient than the proposed method, produces a very sub-optimal result compared to THESIS. Overall, the THESIS approach is more suitable for application in UDHNs because of its combined performance efficiency. In future, we intend to address the problem of insufficient readio resources for traffic offloading at the MBSs by replacing them with unmanned aerial vehicle based BSs~(UAV-BSs) in order to investigate the effect of increased radio resources for traffic offloading and flexibility of BS deployment on the energy consumption of the network.

\section*{Acknowledgement}
This work was supported by the Tertiary Education Trust Fund~(TETFund) of the Federal Republic of Nigeria.
\bibliographystyle{IEEEtran}
\bibliography{ref}
\end{document}